\documentclass[11pt, a4paper]{article}
\usepackage{amssymb,amsbsy,amsmath,amsfonts,amssymb,amscd,colordvi}
\usepackage{a4wide}
\usepackage{times}
\usepackage{comment}
\usepackage{euscript}
\usepackage{verbatim}
\usepackage{fancyhdr}
\usepackage[usenames,dvipsnames]{color}
\usepackage{subcaption}
\usepackage{tikz}
\usepackage{float}
\usepackage{color}
\usepackage{graphicx}
\usepackage{epstopdf}
\usepackage[T1]{fontenc}
\usepackage{cite}
\usepackage{epsfig}
\usepackage{ulem}
\def \trait (#1) (#2) (#3){\vrule width #1pt height #2pt depth #3pt}
\def \fin{\hfill
	\trait (0.1) (5) (0)
	\trait (5) (0.1) (0)
	\kern-5pt
	\trait (5) (5) (-4.9)
	\trait (0.1) (5) (0)
\medskip}
\def \ve{\varepsilon} 

\def\mx{\mathsf m}
\def\fr{\mathsf f}
\def \bsxi{\boldsymbol\xi}

\def \bsvt{\boldsymbol\vartheta}

\def \to{\rightarrow}

\def \eqdef{\stackrel {\rm def} {=}}

\def \fr{\mathsf f}

\newtheorem{lemma}{Lemma}

\newtheorem{remark}{Remark}

\title{New non-equilibrium matrix imbibition equation for Kondaurov's double porosity model}

\author{Andrey Konyukhov$^{\,\dagger,\ddagger}$ and Leonid Pankratov$^{\,\dagger}$}

\begin{document}

\renewcommand{\theequation}{\thesection .\arabic{equation}}

\renewcommand{\baselinestretch}{1.2}

\maketitle

\begin{small}
\noindent
$^{\dagger}$Laboratory of Fluid Dynamics and Seismic, Moscow Institute of Physics and Technology,
9 Institutskiy per., Dolgoprudny, Moscow Region, 141700, Russian Federation.

\noindent
$^{\ddagger}$ Joint Institute for High Temperatures of the Russian Academy of Sciences, Izborskaya
13 Bldg, 2, Moscow, 125412, Russian Federation.

E--mail: {\tt leonid.pankratov@univ-pau.fr} and {\tt konyukhov$\_$av@mail.ru}
\end{small}

\begin{abstract}
The paper deals with the global Kondaurov double porosity model
describing a non-equilibrium two-phase immiscible flow in fractured-porous reservoirs
when non-equilibrium phenomena occur in the matrix blocks, only. It is shown that the
homogenized model can be represented as usual equations of two-phase incompressible
immiscible flow, except for the addition of two source terms calculated by a solution
to a local problem which is a boundary value problem for a non-equilibrium
imbibition equation given in terms of the real saturation and a non-equilibrium parameter.
\end{abstract}

\noindent
{\bf Keywords:} homogenization; double porosity media; immiscible; two-phase flow; non-equilibrium model.
\bigskip

\noindent
{\bf AMS Subject Classifications:} 35B27, 35K65, 35Q35, 74Q10, 74Q15.

\section{Introduction}
\label{intro}

The homogenized Kondaurov double porosity type model (see \cite{ak-lp-AA}) describing a
non-equilibrium two-phase flow of immiscible incompressible fluids in fractured-porous reservoirs is considered.
Two-phase flow in porous media is important to many practical problems, including those in
petroleum reservoir engineering, soil science etc. The modeling and
numerical simulation of two-phase flow in porous media represents an important key in the design
of cost-efficient, safe and suitable clean-up tools. It can reduce the number of laboratory
and field experiments, help to identify the significant mechanisms, optimize existing strategies
and give indications of possible risks. In the existing physical and mathematical literature
the authors are dealing mainly with the equilibrium models. However, the experimental studies
have invalidated this kind of models (see, e.g., \cite{bot}).
The model considered in this paper corresponds physically to a non-equilibrium immiscible
incompressible two-phase flow through fractured porous media. Notice that the crucial feature
of a porous medium, saturated with immiscible fluids, is the fact that the process
depends on the rate and direction of the change of state. The most well-known and often discussed phenomena
of this type are the relaxation of capillary pressure, the "capillary pressure-saturation" hysteresis
curve, and the dependence of the phase permeabilities and the value of the capillary sticking on
the rate and direction of a change in the saturation. The generally accepted explanation of
these phenomena is the non-equilibrium of the joint motion of the fluids (see, e.g., \cite{coussy}).

The homogenization of multiphase flow through heterogeneous porous media
as well as the numerical simulation of this physical process has been
a problem of interest for many years and many methods have been developed. There is an
extensive literature on this subject. Here we refer to the monographs \cite{hor,panf}
as well as to \cite{sal-bru-2010a,sal-bru,salimi}. A recent review of the mathematical
homogenization methods developed for two-phase flow in porous media can be viewed in
\cite{our-siam,m2as}. It is important to notice that the
microscopic models of the multiphase flow in porous media considered in all these
works are equilibrium even if the homogenization process for single- and multiphase flows in
double porosity media leads to appearing of an additional source term which exhibits the global
non-equilibrium behavior of the model (see, e.g., \cite{m2as,blm,panf,yeh2}). In addition,
it is shown in \cite{AMPP,Bourgeat-Panfilov} that the homogenization procedure leads to
appearing of the non-equilibrium capillary pressure in the global model. However,
there are few papers dealing with the homogenization of non-equilibrium two-phase
flows in porous media. Here we refer to \cite{salimi}, where the authors deal
with the upscaling of such flows in vertically fractured oil reservoirs. The homogenization
process is carried out for Barenblatt's and Hassanizadeh's flow models
(see, e.g., \cite{bar-patz-sil,hass,das-book}). Concerning the rigorous mathematical studies
in the domain of non-equilibrium two-phase flows, we also observe only few
papers on the subject dealing mainly with the existence and uniqueness problems (see, e.g.,
\cite{pop,KRS}).

In this paper we study an immiscible non-equilibrium
two-phase flow in double porosity media in the framework of the thermodynamically consistent
Kondaurov model \cite{Kond,a-kon} which is, in fact, an integro-differential one
due to the fact that the mobility functions and the capillary pressure depend on Kondaurov's
non-equilibrium parameter which satisfies a kinetic equation with respect to the real saturation
(see Section \ref{phys-mod} below). The detailed comparison
of the Kondaurov model and Barenblatt's and Hassanizadeh's
non-equilibrium flow models is done in \cite{a-kon,ak-lp-AA}. Here we focus our attention on
the homogenized non-equilibrium double porosity type model obtained recently in \cite{ak-lp-AA}.
This model has a rather complicated form in vue of the numerical simulation. From the other hand,
we know that the numerical methods are very sensitive to the choice of the governing equations form.
Then the aim of the present paper is to find a more simple form for the local problem involved in the
model. Namely, we will show that the homogenized problem can be represented as usual equations of
two-phase incompressible, immiscible flow, with two source terms calculated by a solution to a local problem
which is a boundary value problem for a {\sl non-equilibrium imbibition equation}. The derivation
of this equation is essentially based on the introduction of a {\sl non-equilibrium global pressure}
which generalizes the notion of the well-known global pressure function (see, e.g.,
\cite{ant-kaz-mon1990,GC-JJ,gals-cras}) widely used in the mathematical analysis of multi-phase flows
in porous media. To our knowledge it is a first attempt of introduction of the non-equilibrium imbibition
equation in the homogenization process.

The rest of the paper is organized as follows. In Section \ref{phys-mod} we present a mathematically
rigorous adimensionalized non-equilibrium Kondaurov model focusing on the correct definitions
of the capillary pressure and mobility functions. In Section \ref{glob-mod-cras}, following the lines
of \cite{ak-lp-AA}, we introduce the global Kondaurov double porosity model. Finally, in
Section \ref{sec-non-eq-imb}, we study the local problem involved in the homogenized model. Introducing
the notion of non-equilibrium global pressure, we reduce the local problem formulated in terms
of phase pressures to a unique non-equilibrium imbibition equation which is an integro-differential
equation with respect to the real saturation. As it shown in Remark \ref{rem-limit-cas} in
Section \ref{sec-non-eq-imb}, the last one is a generalization
of the well-known imbibition equation appearing in the homogenization of the two-phase double porosity
models (see, e.g., \cite{Jurak} and the references herein). The paper is completed by the concluding remarks.

\section{Adimensionalized non-equilibrium Kondaurov model}
\label{phys-mod}

In this section we introduce the adimensionalized non-equilibrium
Kondaurov flow model proposed in \cite{Kond} and then developed
in \cite{a-kon}. More recently it was discussed in \cite{ak-lp-AA}.
The equations of the model read:
\begin{equation}
\label{debut1-phys-nonD}
\Phi\, \frac{\partial S_{\kappa}}{\partial t} + {\rm div}\, {\bf W}_{\kappa}  = 0,
\,\,\, {\rm where\,\, the\,\, fluxes\,\, are\,\, defined\,\, by:}\,\,
{\bf W}_{\kappa} = -\frac{K f_{\kappa}(S_\kappa,\bsxi)}{\mu_{\kappa}}
\nabla p_{\kappa} \,\, (\kappa = w, n).
\end{equation}
Here the subscripts $w, n$ denote the wetting and non-wetting fluids; $S_\kappa$ is the saturation of
the corresponding fluid; $0 < \Phi < 1$ is the porosity function; $K$ is the absolute permeability tensor;
$p_\kappa$ is the pressure of the wetting (non-wetting) fluid;
$f_{\kappa} = f_{\kappa}(S_\kappa,\bsxi)$ stands for the relative permeability of the wetting (non-wetting)
fluid defined by:
\begin{equation}
f_{w}(S_w, \bsxi) = f_{w}^{\rm e}\left(2\,S_w + \beta\,\bsxi/\alpha - 1\right)
\quad {\rm and} \quad
f_{n}(S_n, \bsxi) = f_{n}^{\rm e}\left(2\,\big[1 - S_w\big] - \beta\,\bsxi/\alpha \right)
\label{eq7-w-n}
\end{equation}
with the superscript "e" denoting the equilibrium relative phase permeabilities in the
Darcy-Muskat law (see, e.g., \cite{GC-JJ}) and $\alpha, \beta > 0$ being constitutive parameters of the
model; $\mu_\kappa$ is the viscosity of the wetting (non-wetting) fluid; finally, $\bsxi$ denotes
the non-equilibrium Kondaurov parameter which satisfies the following kinetic equation:
\begin{equation}
\frac{{\partial \bsxi }}{{\partial t}} = \frac{1}{\tau}\,\Lambda(S_{w}, \bsxi)
\quad {\rm with} \,\,\,
\Lambda (S_n,\bsxi) \eqdef \frac{\alpha}{\beta}\,[1 - S_w] - \bsxi.
\label{eq9e}
\end{equation}
Here $\tau > 0$ is the relaxation time. The model is completed as follows. By the definition
of saturations, one has $S_{w} + S_{n} = 1$ with $S_{w}, S_{n} \geqslant 0$. Then the curvature
of the contact surface between the two fluids links the jump of pressure of two phases to the saturation
by the capillary pressure law: $P_{\rm c}(S_w,\bsxi) = p_{n} - p_{w}$,
where (see, e.g., \cite{ak-lp-AA}) the capillary pressure function has the form:
\begin{equation}
\Phi P_{\rm c}(S_w,\bsxi) \eqdef \gamma  + M\, [1 - S_w] - \alpha\, \bsxi.
\label{eq8}
\end{equation}
Here $M, \gamma > 0$ are constitutive parameters of the model. Finally, we introduce the
mobility functions $\lambda_\kappa$ which will be widely used below. They are defined as:
$\lambda_\kappa(S_\kappa, \bsxi) \eqdef f_{\kappa}(S_\kappa, \bsxi)/\mu_\kappa$ ($\kappa =
w, n$).

Now we discuss in more details the definitions and the properties of the capillary pressure
and the mobility functions. We also formulate the conditions on the
constitutive parameters of the model. We start the analysis by establishing the explicit
dependence of the non-equilibrium parameter on the wetting saturation function.
Namely, denoting $S\eqdef S_w$, one can easily show that
\begin{equation}
\label{xi-ell-5-new}
\bsxi = \bsxi^{\rm init}(x)\, e^{-{t}/{\tau}} +
\frac{\alpha}{\tau\beta}\,
\int_0^t  e^{{(\varsigma-t)}/{\tau}}\,\big(1 - S(x, \varsigma)\big)\, d\varsigma
\quad {\rm with}\,\,\bsxi^{\rm init}(x) \eqdef \bsxi(x, 0) > 0.
\end{equation}
Consider the capillary pressure function. The initial boundary value problem
for the two-phase filtration is well posed if only if the capillary pressure function $P_{\rm c}$
is a decreasing function of the saturation $S$. In order to prove this fact we often
deal with the derivative of the parameter $\bsxi$ with respect to $S$. This derivative involves the function
$\bsxi^\prime_S \eqdef \frac{\partial \bsxi}{\partial S}(x, 0)$. From now on, for the sake
of definiteness, we assume that $\bsxi^\prime_S = \bsxi^\prime_S(x, 0) \geqslant 0$ in $\Omega$,
where $\Omega$ is our reservoir of interest. We have the following result.
\medskip

\begin{lemma}
\label{decreas-lem}
Let the function $\bsxi^\prime_S \geqslant 0$ satisfy the bound $\max_{x\in\Omega}
\bsxi^\prime_S(x, 0) < + \infty$ in $\Omega$ and let $M, \alpha, \beta$ be such that
$M > 2\,\alpha^2/\beta$. Then the function $P_{\rm c}$ is a positive decreasing function of $S$.
\end{lemma}
\medskip

The proof of the lemma is based on the application of the kinetic equation (\ref{eq9e}).
\medskip

Now we turn to the mathematically rigorous definition and the properties of the mobility
functions $\lambda_w, \lambda_n$ in the non-equilibrium case. Let us recall that for an
equilibrium two-phase flow in porous medium (see, e.g., \cite{our-siam,AMPP}, and the
references therein) the standard assumptions on the mobility functions are:
$0 \leqslant \lambda^{\rm e}_w(S), \lambda^{\rm e}_{n}(1 - S) \leqslant 1$ for
$S \in [0, 1]$ and $\lambda^{\rm e}_w(S=0) = 0$, $\lambda^{\rm e}_w(S=1) = 1$ and
$\lambda^{\rm e}_n(S=0) = 1$, $\lambda^{\rm e}_n(S=1) = 0$. Here $S$ stands
for the wetting phase saturation in the equilibrium case. Our goal now is to
establish similar properties of the mobility functions which depend, in the non-equilibrium
case, both on the real saturation $S$ and the non-equilibrium parameter $\bsxi$.
To this end, it is natural to introduce a new non-equilibrium parameter $\bsvt$ given by:
\begin{equation}
\label{nsec-6}
\bsvt \eqdef 2\,S + \beta\,\bsxi/\alpha - 1
\end{equation}
and to consider the properties of the mobility functions in terms of
this parameter. As functions of $\bsvt$, the mobilities become:
$\lambda_{w}(S, \bsxi) = \lambda^{\rm e}_{w}(\bsvt)$ and
$\lambda_{n}(S, \bsxi) = \lambda^{\rm e}_{n}(1 - \bsvt)$.
We have:
\medskip

\begin{lemma}
\label{0,1-lemma}
Let $\bsvt$ be the parameter defined in (\ref{nsec-6}). Assume that
$0 < \max_\Omega \frac{\beta}{\alpha}\,\bsxi^{\rm init}(x) < 1$ in $\Omega$.
Then we have: {\bf (i)} There are the values of the saturation $S$, denoted by $S_{\bsvt = 0}$ and
$S_{\bsvt = 1}$, such that
\begin{equation}
\label{bseta=0-eq1}
\bsvt = 0 \,\,\, {\rm for} \,\,\, S_{\bsvt = 0} \eqdef\,
\frac{e^{-t/(2\tau)}}{2}\,\left(1 - \frac{\beta}{\alpha}\,\bsxi^{\rm init}\right)
\qquad {\rm and} \qquad
\bsvt = 1 \,\,\, {\rm for} \,\,\, S_{\bsvt = 1} \eqdef\,
1 - \frac{\beta}{2\alpha}\bsxi^{\rm init}(x)\, e^{-t/(2\tau)}.
\end{equation}
{\bf (ii)} The values $S_{\bsvt = 0}, S_{\bsvt = 1}$ are such that
$0 < S_{\bsvt = 0} < S_{\bsvt = 1} < 1$ and $S_{\bsvt = 0} \to 0$,
$S_{\bsvt = 1} \to 1$ as $t \to + \infty$.
\end{lemma}
\medskip

\noindent In order to prove the lemma, one have to solve a Volterra nonhomogeneous
equation coming from the representation (\ref{xi-ell-5-new}) of the non-equilibrium parameter $\bsxi$.

Now, let us study the dependance of the parameter $\bsvt$ on the saturation $S$.
We have:
\smallskip

\begin{lemma}
\label{bsvt-satur-lemma}
Let $\bsvt$ be the parameter defined by (\ref{nsec-6}). Then $\bsvt$ is an increasing function of $S$.
\end{lemma}
\smallskip

\noindent The proof of the lemma is based on the positiveness of the function $\bsxi^\prime_S$.
\smallskip

Thus we conclude that with the following assumptions on the constitutive parameters:
\begin{equation}
\label{bsvt-eq-5}
\bsxi^\prime_S(x, 0) \geqslant 0, \quad
M > 2\,\alpha^2/\beta, \quad {\rm and} \quad
0 < \max_\Omega \left(\beta\,\bsxi^{\rm init}(x)/\alpha\right) < 1 \quad {\rm in\,\,} \Omega
\end{equation}
we have that: {\bf (i)} The capillary pressure is a decreasing function of the saturation $S$.
{\bf (ii)} The parameter $\bsvt$ equals 0 and 1 for $S_{\bsvt = 0}$ and
$S_{\bsvt = 1}$ given by (\ref{bseta=0-eq1}). {\bf (iii)} The parameter $\bsvt$ is an
increasing function of $S$.

Now let us explain how do we understand the mobility functions $\lambda_w, \lambda_n$
in our further analysis. We set:
\begin{equation}
\label{def-llla-w}
\lambda_w(S, \bsxi) :=
\left\{
\begin{array}[c]{ll}
1, \quad {\rm when} \,\, S > S_{\bsvt = 1};\\
\lambda^{\rm e}_w(\bsvt) \quad {\rm when} \,\,
S \in I_{\bsvt};\\
0, \quad {\rm when} \,\, S < S_{\bsvt = 0}
\end{array}
\right.
\quad {\rm and} \quad
\lambda_n(S, \bsxi) :=
\left\{
\begin{array}[c]{ll}
1, \quad {\rm when} \,\, S < S_{\bsvt = 0};\\
\lambda^{\rm e}_n(1 - \bsvt) \quad {\rm when} \,\,
S \in I_{\bsvt}; \\
0, \quad {\rm when} \,\, S > S_{\bsvt = 1},
\end{array}
\right.
\end{equation}
where $I_{\bsvt} \,\eqdef\, [S_{\bsvt = 0}, S_{\bsvt = 1}]$ stands for our interval of interest.

\section{The global Kondaurov double porosity model}
\label{glob-mod-cras}

In this section we formulate the mesoscopic flow equations of the Kondaurov model
and then introduce the homogenized model obtained earlier in \cite{ak-lp-AA}. We consider
a reservoir $\Omega \subset \mathbb{R}^d$ ($d = 2, 3$)
which is assumed to be a bounded, connected domain with a periodic structure.
More precisely, we will scale this periodic structure by a
parameter $\ve$ which represents the ratio of the cell size to
the whole region $\Omega$  and we assume that $\ve \downarrow 0$. Let $Y \eqdef (0, 1)^d$
be a basic cell of a fractured porous medium. We assume that $Y$ is made
up of two homogeneous porous media $Y_\mx$ and $Y_\fr$ corresponding to the parties
of the mesoscopic domain occupied by the matrix block and the fracture, respectively.
Thus $Y = Y_\mx \cup Y_\fr \cup \Gamma_{\fr\mx}$, where $\Gamma_{\fr\mx}$ denotes
the interface between the two media. Let $\Omega^\ve_\ell$ with
$\ell = "\fr"$ or $"\mx"$ denotes the open set corresponding to the porous medium with index
$\ell$. Then $\Omega = \Omega^\ve_\mx \cup \Gamma^\ve_{\fr\mx} \cup \Omega^\ve_\fr$,
where $\Gamma^\ve_{\fr\mx} \eqdef \partial \Omega^\ve_\fr \cap \partial \Omega^\ve_\mx \cap \Omega$
and the subscripts $"\mx"$, $"\fr"$ refer to the matrix and fracture, respectively.

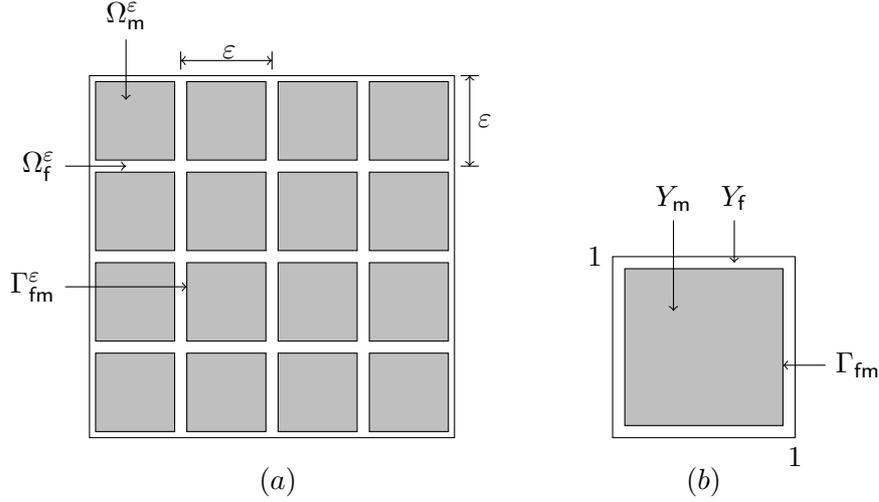
\begin{figure}[H]

\begin{center}

\begin{tikzpicture}[scale=0.8]
\def\rectanglepath{-- ++(1.3cm,0cm)  -- ++(0cm,1.3cm) -- ++(-1.3cm,0cm) -- cycle}

\filldraw[fill=white!30] (-0.1,-0.1) -- (5.9,-0.1)
                                         --  (5.9, 5.9) -- (-0.1, 5.9) -- cycle;

\filldraw[fill=lightgray] (0,  0) \rectanglepath;
\filldraw[fill=lightgray] (1.5,0) \rectanglepath;
\filldraw[fill=lightgray] (3,  0) \rectanglepath;
\filldraw[fill=lightgray] (4.5,0) \rectanglepath;

\filldraw[fill=lightgray] (0,  1.5) \rectanglepath;
\filldraw[fill=lightgray] (1.5,1.5) \rectanglepath;
\filldraw[fill=lightgray] (3,  1.5) \rectanglepath;
\filldraw[fill=lightgray] (4.5,1.5) \rectanglepath;

\filldraw[fill=lightgray] (0,  3.0) \rectanglepath;
\filldraw[fill=lightgray] (1.5,3.0) \rectanglepath;
\filldraw[fill=lightgray] (3,  3.0) \rectanglepath;
\filldraw[fill=lightgray] (4.5,3.0) \rectanglepath;

\filldraw[fill=lightgray] (0,  4.5) \rectanglepath;
\filldraw[fill=lightgray] (1.5,4.5) \rectanglepath;
\filldraw[fill=lightgray] (3,  4.5) \rectanglepath;
\filldraw[fill=lightgray] (4.5,4.5) \rectanglepath;

\draw[<-] (0.5,5.5) -- (0.5,6.5) node[anchor=south] {$\Omega_\mx^{\varepsilon}$};
\draw[<-] (0.5,4.4) -- (-0.5,4.4) node[anchor=east] {$\Omega_\fr^{\varepsilon}$};
\draw[<-] (1.5,2.4) -- (-0.5,2.4) node[anchor=east] {$\Gamma^\ve_{\fr\mx}$};

\draw (1.4,6.0)--(1.4,6.3);
\draw (2.9,6.0)--(2.9,6.3);
\draw[<->] (1.4,6.15)--(2.9,6.15);
\draw (2.2,6.3) node {$\varepsilon$};

\draw (6.0,5.9)--(6.3,5.9);
\draw (6.0,4.4)--(6.3,4.4);
\draw[<->] (6.15,4.4)--(6.15,5.9);
\draw (6.4,5.15) node {$\varepsilon$};



\node [below=0.5cm]
at (3.0,0.2)
{$(a)$};


\def\rectanglepath{-- ++(1.3cm,0cm)  -- ++(0cm,1.3cm) -- ++(-1.3cm,0cm) -- cycle}

\draw[fill=white!30] (8.5,-0.1) -- (11.5,-0.1) -- (11.5,2.9) -- (8.5,2.9) -- cycle;
\draw[fill=lightgray]   (8.7,0.1) -- (11.3,0.1) -- (11.3,2.7) -- (8.7,2.7) -- cycle;

\draw (11.5,-0.1) node[anchor=north] {$1$};
\draw (8.5,2.9) node[anchor=east] {$1$};

\draw[<-] (9.5,2.0) -- (9.5,3.5) node[anchor=south] {$Y_\mx$};
\draw[<-] (10.5 ,2.8) -- (10.5,3.5) node[anchor=south] {$Y_\fr$};
\draw[<-] (11.3 ,1.1) -- (12 ,1.1) node[anchor=west] {$\Gamma_{\fr\mx}$};



\node [below=0.5cm]
at (10.0,0.2) {$(b)$};

\end{tikzpicture}

\end{center}

\caption{(a) The domain $\Omega$ with the mesostructure. \quad (b) The reference cell $Y$.}
\label{fig:ref}

\end{figure}

Before describing the equations of the model (\ref{debut1-phys-nonD}) for the nonhomogeneous
porous medium $\Omega$ with the periodic structure, we give the corresponding notation and
also define the porosity function and the global permeability tensor adopted to the
double porosity medium $\Omega$. We have: $\Phi^\ve(x) = \Phi(\frac{x}{\ve})$ is the
porosity of the reservoir $\Omega$. The function $\Phi^\ve$ is a $Y$-periodic defined by:
$\Phi^\ve(x) \eqdef \Phi_\fr {\bf 1}^\ve_\fr(x) + \Phi_\mx {\bf 1}^\ve_\mx(x)$,
where ${\bf 1}^\ve_\fr, {\bf 1}^\ve_\mx$ are the characteristic functions of the media
$\Omega^\ve_\fr$, $\Omega^\ve_\mx$, respectively, and where the constants $0 < \Phi_\fr,
\Phi_\mx < 1$ do not depend on $\ve$; $K^\ve(x) = K(\frac{x}{\ve})$ is the absolute
permeability tensor of $\Omega$ it is defined by: $K^\ve(x) \eqdef K_\fr\, {\bf 1}^\ve_\fr(x) +
\ve^2 K_\mx\,{\bf 1}^\ve_\mx$, where $0 < K_\fr$, $K_\mx < +\infty$ are positive constants
that do not depend on $\ve$; $S^\ve_{\ell, w} = S^\ve_{\ell, w}(x, t)$,
$S^\ve_{\ell, n} = S^\ve_{\ell, n}(x, t)$ are the saturations of wetting and nonwetting fluids in
$\Omega^\ve_\ell$, respectively; $p^\ve_{\ell, w} = p^\ve_{\ell, w}(x,t)$, $p^\ve_{\ell, n} =
p^\ve_{\ell, n}(x,t)$ are the pressures of wetting and nonwetting fluids in
$\Omega^\ve_\ell$, respectively; $\bsxi^\ve_{\ell} = \bsxi^\ve_{\ell}(x,t)$ is the non-equilibrium parameter
in the medium $\Omega^\ve_\ell$; $\lambda_{\ell,w}, \lambda_{\ell,n}$ are the mobilities
of wetting and nonwetting fluids in $\Omega^\ve_\ell$, respectively; $\tau_\ell$ is the relaxation
time in $\Omega^\ve_\ell$; $\alpha_\ell, \beta_\ell, \gamma_\ell, M_\ell > 0$ denote the
constitutive parameters in $\Omega^\ve_\ell$ which do not depend on $\ve$.
Denoting $S^\ve_\ell \eqdef S^\ve_{\ell,w}$, we obtain the following flow equations:
\begin{equation}
\label{debut2}
\left\{
\begin{array}[c]{ll}
\displaystyle
\Phi^\ve(x) \frac{\partial {\mathsf S}^\ve}{\partial t} - {\rm div}\, \left\{K^\ve(x)
\lambda_{w}\left(\frac{x}{\ve}, {\mathsf S}^\ve, {\boldsymbol\bsxi}^\ve \right)
\nabla {\mathsf p}^\ve_{w} \right\} = 0 \quad {\rm in}\,\, \Omega_{T}; \\[3mm]
\displaystyle
-\Phi^\ve(x) \frac{\partial {\mathsf S}^\ve}{\partial t} -
{\rm div}\, \left\{K^\ve(x) \lambda_{n}\left(\frac{x}{\ve}, {\mathsf S}^\ve,
{\boldsymbol\bsxi}^\ve \right)
\nabla {\mathsf p}_{n} \right\} =  0 \quad {\rm in}\,\, \Omega_{T}; \\[3mm]
\displaystyle
P^\ve_{c}\left(\frac{x}{\ve}, {\mathsf S}^\ve, {\boldsymbol\bsxi}^\ve\right) =
{\mathsf p}^\ve_{n} - {\mathsf p}^\ve_{w}
\,\,{\rm with}\,\, \Phi^\ve(x) P^\ve_{c}\left(\frac{x}{\ve},
{\mathsf S}^\ve, {\boldsymbol\bsxi}^\ve\right) \eqdef
\gamma^\ve(x) + M^\ve(x)\, [1 - {\mathsf S}^\ve] -
\alpha^\ve(x)\, {\boldsymbol\bsxi}^\ve
,\\
\end{array}
\right.
\end{equation}
where $\Omega_T \eqdef \Omega \times (0,T)$ ($T > 0$ is fixed);
the mobilities $\lambda_{\ell,w}, \lambda_{\ell,n}$ are defined (in accordance with
(\ref{def-llla-w})) by:
\begin{equation}
\label{mob-w-n}
\lambda_{\ell,w}(S^\ve_{\ell}, \bsxi^\ve_{\ell}) =
\lambda_{\ell,w}\!\left(2\,S^\ve_{\ell} - 1 + \beta_\ell\,\bsxi^\ve_{\ell}/\alpha_\ell\right)
\quad {\rm and} \quad
\lambda_{\ell,n}(S^\ve_{\ell}, \bsxi^\ve_{\ell}) = \lambda_{\ell,n}\!\left(2\,\big[1 - S^\ve_{\ell}\big] -
\beta_\ell\,\bsxi^\ve_{\ell}/\alpha_\ell\right)
\end{equation}
and each function
$u^\ve := {\mathsf S}^\ve, {\mathsf p}^\ve_{w}, {\mathsf p}^\ve_{n}, {\boldsymbol\bsxi}^\ve$
as well as the piece-wise constant functions $\Phi^\ve, K^\ve, \gamma^\ve, M^\ve, \alpha^\ve$ are defined as:
$u^\ve \eqdef u_\fr^\ve {\bf 1}^\ve_\fr(x) + u_\mx^\ve {\bf 1}^\ve_\mx(x)$. The system (\ref{debut2})
is completed by the corresponding interface and initial conditions which are omitted here
for the sake of brevity (for more details see \cite{ak-lp-AA}).

Now we introduce the global non-equilibrium Kondaurov flow model obtained by the
method of two-scale asymptotic expansions (see, e.g., \cite{BAKH-PAN,BLP,Bourgeat-Panfilov,SP})
in Section 4.2 of \cite{ak-lp-AA}.
Here we also restrict ourselves to a special case of the homogenized model.
Namely, as in \cite{salimi}
we consider the non-equilibrium effects for the matrix part only and not for the fracture
system which is related to the fact that the non-equilibrium effects for fractures,
due to their high permeabilities and, consequently, low capillary forces, are negligible.
First, we introduce the notation: $S$, $P_w$, $P_n$ denote the homogenized wetting
liquid saturation, the wetting and nonwetting liquid pressures, respectively;
$\Phi^\star$ denotes the effective porosity and is given by:
$\Phi^\star \eqdef \Phi_{\fr}\,|Y_\fr|/|Y_\mx|$,
where $|Y_\ell|$ is the measure of the set $Y_\ell$ ($\ell = \fr, \mx$);
$\mathbb{K}^\star$ is the homogenized tensor with the entries
\begin{equation}
\label{H-2}
\mathbb{K}^{\star}_{ij} \eqdef \frac{K_\fr}{|Y_\mx|}\, \int_{Y_\fr}\,
\left[\nabla_y \zeta_i + \vec e_i \right]\, \left[\nabla_y \zeta_j + \vec e_j \right]\, dy,
\,\, {\rm where}\,\, \zeta_j\,\, {\rm satisfies:}\,\, \left\{
\begin{array}[c]{ll}
- \Delta_y\, \zeta_j = 0 \,\, {\rm in} \,\, Y_{\fr}; \\
\nabla_y \zeta_j \cdot \vec \nu_y = - \vec e_j \cdot \vec \nu_y
\,\, {\rm on} \,\, \Gamma_{\fr\mx}\\
y \mapsto \zeta_j(y)\quad Y-{\rm periodic}. \\
\end{array}
\right.
\end{equation}
Then the homogenized system has the form:
\begin{equation}
\label{F-present-1-sal}
\left\{
\begin{array}[c]{ll}
\displaystyle
\Phi^\star\, \frac{\partial S}{\partial t}
- {\rm div}_x\, \bigg\{\mathbb{K}^\star\, \lambda_{\,\fr,w}(S) \nabla P_w \bigg\}
= {\EuScript Q}_w \quad {\rm in} \,\, \Omega_T;\\[3mm]
\displaystyle
- \Phi^\star\, \frac{\partial S}{\partial t}
- {\rm div}_x\, \bigg\{\mathbb{K}^\star\, \lambda_{\,\fr,n}(1 - S)
\nabla P_n \bigg\}
= {\EuScript Q}_n \quad {\rm in} \,\, \Omega_T;\\[3mm]
\displaystyle
P_{c}(S) = P_{n} - P_{w}\,\, {\rm with}\,\, \Phi_\fr\, P_{c}(S) \eqdef
{\mathsf a}_{\fr,1}\,S + {\mathsf a}_{\fr,3} \quad {\rm in}\,\, \Omega_{T},
\end{array}
\right.
\end{equation}
where the constants ${\mathsf a}_{\fr,j}$ ($j = 1, 2, 3$) in vue of condition
(\ref{bsvt-eq-5}) are defined as:
\begin{equation}
\label{const-aaa}
{\mathsf a}_{\ell,1} \eqdef -\left(M_\ell - \alpha^2_\ell/\beta_\ell\right) < 0,
\quad {\mathsf a}_{\ell,2} \eqdef
\tau_\ell\,\left(M_\ell - 2\,\alpha^2_\ell/\beta_\ell \right) > 0,
\,\, {\mathsf a}_{\ell,3} \eqdef \gamma_\ell + M_\ell -
\alpha^2_\ell/\beta_\ell > 0 \,\, (\ell = \fr, \mx).
\end{equation}

\begin{remark}
Notice that the functions $S$, $P_w$, $P_n$ appearing in (\ref{F-present-1-sal})
are, in fact, zero order terms in the asymptotic expansions for the saturations
$S^\ve_\fr$, and phase pressures $p^\ve_{\fr,w}$, $p^\ve_{\fr,n}$ in the fracture
domain $\Omega^\ve_\fr$ (for more details see formulae (3.3)-(3.4) and the beginning
of Section 3.2 in \cite{ak-lp-AA}). In a similar way, we introduce below the functions
$s, p_w, p_n$ in (\ref{F-present-2}) for the matrix block.
\end{remark}

For almost all point $x \in \Omega$, the equations for flow in a matrix block are given by:
\begin{equation}
\label{F-present-2}
\left\{
\begin{array}[c]{ll}
\displaystyle
\Phi_\mx\, \frac{\partial s}{\partial t} -
{\rm div}_y\, \bigg\{K_\mx\, \lambda_{\,\mx,w}(\bsvt_\mx) \nabla_y p_{w} \bigg\}
= 0 \quad {\rm in} \,\, Y_\mx\times\Omega_T; \\[3mm]
\displaystyle
- \Phi_\mx\, \frac{\partial s}{\partial t}
-
{\rm div}_y\, \bigg\{K_\mx\, \lambda_{\,\mx,n}(1 - \bsvt_\mx)
\nabla_y p_{n} \bigg\} = 0 \quad {\rm in} \,\, Y_\mx\times\Omega_T;\\[5mm]
\displaystyle
p_c\left(\bsvt_\mx, \frac{\partial \bsvt_\mx}{\partial t}\right) = p_n - p_w
\,\,{\rm with} \,\,\, \Phi_\mx\, p_c\left(\bsvt_\mx,
\frac{\partial \bsvt_\mx}{\partial t}\right)
\eqdef {\mathsf a}_{\mx,1}\,\bsvt_\mx + {\mathsf a}_{\mx,2}\,
\frac{\partial \bsvt_\mx}{\partial t} + {\mathsf a}_{\mx,3}
;\\[4mm]
p_{w}(x, y, t) = P_{w}(x, t) \quad {\rm and} \quad p_{n}(x, y, t) = P_{n}(x, t)
\quad {\rm on} \,\, \Gamma_{\fr\mx} \times \Omega_T. \\[1mm]
\end{array}
\right.
\end{equation}
Here we make use of the following notation: $s$, $p_w$, $p_n$
denote the wetting liquid saturation, the wetting and nonwetting liquid pressures in
the matrix block $Y_\mx$, respectively; $\bsxi_\mx$ denotes the local non-equilibrium parameter
in the matrix block $Y_\mx$, it is defined as the solution to the following equation:
\begin{equation}
\label{H-xi-mx}
\frac{\partial \bsxi_\mx}{\partial t} = \frac{1}{\tau_\mx}\, \Lambda(s, \bsxi_\mx)
\quad {\rm with \,\,}
\Lambda(s, \bsxi_\mx) \eqdef \frac{\alpha_\mx}{\beta_\mx}\,\big[1 - s \big] - \bsxi_\mx;
\end{equation}
the parameter $\bsvt_\mx$ is defined by: $\bsvt_\mx \eqdef 2\,s + \beta_\mx\,\bsxi_\mx/\alpha_\mx - 1$.
For any $x \in \Omega$ and $t > 0$, the matrix-fracture sources are given by:
\begin{equation}
\label{F-present-4}
{\EuScript Q}_w \eqdef - \frac{\Phi_\mx}{|Y_\mx|}\, \int_{Y_\mx}
\frac{\partial s}{\partial t}(x, y, t) \,dy = - {\EuScript Q}_n.
\end{equation}

\begin{remark}
\label{remark-on-hom-sys}
Notice that in the case of the equilibrium flow, from (\ref{H-xi-mx}), we have that
$\bsxi_{\mx} = \frac{\alpha_\mx}{\beta_\mx}\,\big[1 - s\big]$ for $\tau_\mx = 0$.
Then the macroscopic model (\ref{F-present-1-sal})-(\ref{F-present-4}) is exactly
(evidently, with a specified capillary pressure) the well known homogenized double
porosity model for the immiscible incompressible two-phase flow in porous media
considered by many authors (see, e.g., \cite{Bourgeat-Panfilov,Jurak,yeh2}
and the references therein).
\end{remark}

\section{Non-equilibrium matrix imbibition equation}
\label{sec-non-eq-imb}

Let us recall that when a porous medium filled with some fluid is brought into contact
with another fluid which preferentially wets the medium, there
is a spontaneous flow of the wetting fluid into the medium and a counterflow of
the resident fluid from the medium. This phenomenon is called imbibition and arises in
physical situations involving multiphase flow systems (see, e.g., \cite{richard}).

The goal of this section is to reduce the local problem (\ref{F-present-2}) formulated
in terms of the phase pressures to a new problem, for a {\sl non-equilibrium imbibition
equation} given in terms of the real saturation $s$ and the parameter $\bsvt_\mx$ which is, in
fact, the functional of $s$. To this end, let us rewrite the capillary
pressure function given in (\ref{F-present-2})$_3$ as follows:
\begin{equation}
\label{F-polu-glob-2}
p_{\,\rm c}\left(\bsvt_\mx, \frac{\partial \bsvt_\mx}{\partial t}\right) = \pi_{\rm c}(\bsvt_\mx) +
\widehat{\mathsf a}_{\mx,2}\,\frac{\partial \bsvt_\mx}{\partial t} \quad {\rm with} \,\,
\pi_c(\bsvt_\mx) \eqdef \widehat{\mathsf a}_{\mx,1}\,\bsvt_\mx + \widehat{\mathsf a}_{\mx,3}
\,\,\, {\rm and} \,\, \widehat{\mathsf a}_{\mx,j} \eqdef {\mathsf a}_{\mx,j}/\Phi_\mx.
\end{equation}

Inspired by \cite{our-siam}, we introduce the notion of {\sl non-equilibrium global pressure}
${\mathsf P}$ which is a generalization of the global pressure function
defined earlier (see, e.g., \cite{ant-kaz-mon1990,GC-JJ,gals-cras}) in the the equilibrium case:
\begin{equation}
\label{F-gp1}
p_{w} \eqdef {\mathsf P} + {\mathsf G}_{w}(\bsvt_\mx)
+ \widehat{\mathsf a}_{\mx,2}\,{\EuScript F}_{w}\left(\bsvt_\mx, \frac{\partial \bsvt_\mx}{\partial t}\right)
\quad {\rm and} \quad
p_{n} \eqdef {\mathsf P} + {\mathsf G}_{n}(\bsvt_\mx) +
\widehat{\mathsf a}_{\mx,2}\,{\EuScript F}_{n}\left(\bsvt_\mx, \frac{\partial \bsvt_\mx}{\partial t}\right),
\end{equation}
where the functions ${\mathsf G}_{w}, {\mathsf G}_{n}, {\EuScript F}_{w}, {\EuScript F}_{n}$
will be specified later using several conditions. Roughly speaking,
these conditions are a consequence of the definition of capillary
pressure function (\ref{F-present-2})$_3$. First, we define the functions ${\mathsf G}_{w},
{\mathsf G}_{n}$. Namely, the function ${\mathsf G}_{n}(\bsvt_\mx)$ we choose in the following way:
\begin{equation}
\label{F-gp3}
{\mathsf G}_{n}(\bsvt_\mx) \eqdef
\int_0^{\,\bsvt_\mx} \frac{\lambda_{\,\mx,w}(\varsigma)}{\lambda_\mx(\varsigma)}\,
\pi_{\rm c}^\prime(\varsigma) \, d\varsigma \quad {\rm with} \,\,
\lambda_\mx(\bsvt_\mx) \eqdef \lambda_{\,\mx,w}(\bsvt_\mx) + \lambda_{\,\mx,n}(\bsvt_\mx).
\end{equation}
From now on, $\lambda_{\,\mx,n}(\bsvt_\mx) := \lambda_{\,\mx,n}(1 - \bsvt_\mx)$ and
$\pi^\prime_{\rm c}$ denotes the derivative of the function $\pi$ with respect to its argument.
Notice that the standard assumption on the function $\lambda_\mx$ (see, e.g. \cite{our-siam}
and the references herein) is that there exists a strictly positive constant $L_0$ such that
$\lambda_\mx(\varsigma) \geqslant L_0 > 0$ in $\varsigma \in [0, 1]$. Now, taking into account that
$\pi_c^\prime(\varsigma) = \widehat{\mathsf a}_{\mx,1}$, where $\widehat{\mathsf a}_{\mx,1} < 0$
(see (\ref{const-aaa})), from (\ref{F-gp3}) we get:
\begin{equation}
\label{F-gp3-der-2}
{\mathsf G}_{n}(\bsvt_\mx) = \widehat{\mathsf a}_{\mx,1}\, \int_0^{\,\bsvt_\mx}
\frac{\lambda_{\,\mx,w}(\varsigma)}{\lambda_\mx(\varsigma)} \, d\varsigma
\quad {\rm with} \,\, \nabla_y {\mathsf G}_{n}(\bsvt_\mx) = \widehat{\mathsf a}_{\mx,1}\,
\frac{\lambda_{\,\mx,w}(\bsvt_\mx)}{\lambda_\mx(\bsvt_\mx)}
\, \nabla_y \bsvt_\mx.
\end{equation}
The function ${\mathsf G}_{w}$ is then defined by
${\mathsf G}_{w}(\bsvt_\mx) \eqdef {\mathsf G}_{n}(\bsvt_\mx) - \pi_{\rm c}(\bsvt_\mx)$. This implies the
following formula for the gradient of the function ${\mathsf G}_{w}$:
\begin{equation}
\label{F-gp4.1++}
\nabla_y {\mathsf G}_{w}(\bsvt_\mx) =
- \frac{\lambda_{\,\mx,n}(\bsvt_\mx)} {\lambda_\mx(\bsvt_\mx)}
\pi_{\rm c}^\prime(\bsvt_\mx)\, \nabla_y \bsvt_\mx =
- \widehat{\mathsf a}_{\mx,1} \frac{\lambda_{\,\mx,n}(\bsvt_\mx)}{\lambda_\mx(\bsvt_\mx)}
\nabla_y \bsvt_\mx.
\end{equation}
We notice that $\lambda_{\,\mx,w}(\bsvt_\mx) \nabla_y {\mathsf G}_{w}(\bsvt_\mx) =
\mathfrak{a}(\bsvt_\mx) \nabla_y \bsvt_\mx$ and
$\lambda_{\,\mx,n}(\bsvt_\mx) \nabla_y {\mathsf G}_{n}(\bsvt_\mx) = - \mathfrak{a}(\bsvt_\mx) \nabla_y \bsvt_\mx$,
where
\begin{equation}
\label{gp+5}
\mathfrak{a}(\bsvt_\mx) \eqdef |\widehat{\mathsf a}_{\mx,1}|\,
\frac{\lambda_{\,\mx,n}(\bsvt_\mx)\,\lambda_{\,\mx,w}(\bsvt_\mx)}{\lambda_\mx(\bsvt_\mx)}.
\end{equation}
Let us introduce the following function:
\begin{equation}
\label{upsi-1}
\mathfrak{b}(\bsvt_\mx) \eqdef \int_0^{\,\bsvt_\mx} \mathfrak{a}(\varsigma)\, d\varsigma
=
|\widehat{\mathsf a}_{\mx,1}|\, \int_0^{\,\bsvt_\mx}
\frac{\lambda_{\,\mx,n}(\varsigma)\,\lambda_{\,\mx,w}(\varsigma)}
{\lambda_\mx(\varsigma)}\, d\varsigma.
\end{equation}
Then taking into account the definition of the function $\mathfrak{b}$ we have:
\begin{equation}
\label{bbb-2-1}
\lambda_{\,\mx,w}(\bsvt_\mx) \nabla_y p_{w}
= \lambda_{\,\mx,w}(\bsvt_\mx) \nabla_y {\mathsf P} + \nabla_y \mathfrak{b}(\bsvt_\mx)
+ \lambda_{\,\mx,w}(\bsvt_\mx)\,
\widehat{\mathsf a}_{\mx,2}\,
\nabla_y{\EuScript F}_{w}\left(\bsvt_\mx, \frac{\partial \bsvt_\mx}{\partial t}\right);
\end{equation}
\begin{equation}
\label{bbb-2-2}
\lambda_{\,\mx,n}(\bsvt_\mx) \nabla_y p_{n} = \lambda_{\,\mx,n}(\bsvt_\mx) \nabla_y {\mathsf P} -
\nabla_y \mathfrak{b}(\bsvt_\mx)
+ \lambda_{\,\mx,n}(\bsvt_\mx)\, \widehat{\mathsf a}_{\mx,2}\,
\nabla_y{\EuScript F}_{n}\left(\bsvt_\mx, \frac{\partial \bsvt_\mx}{\partial t}\right).
\end{equation}
Now, we turn to the functions ${\EuScript F}_{w}, {\EuScript F}_{n}$.
The relation (\ref{F-present-2})$_3$ along with the previous assumptions
on the functions ${\mathsf G}_{w}, {\mathsf G}_{n}$ leads to the following condition:
\begin{equation}
\label{F-ass-on-ff+1}
{\EuScript F}_{n}\left(\bsvt_\mx, \frac{\partial \bsvt_\mx}{\partial t}\right)
-
{\EuScript F}_{w}\left(\bsvt_\mx, \frac{\partial \bsvt_\mx}{\partial t}\right)
=
\frac{\partial \bsvt_\mx}{\partial t}.
\end{equation}
Let us rewrite (\ref{F-present-2}) in terms of the non-equilibrium global pressure ${\mathsf P}$,
saturation $s$, and the non-equilibrium parameter $\bsvt_\mx$. From (\ref{bbb-2-1}), (\ref{bbb-2-2}),
we get:
\begin{equation}
\label{fanta-1}
\Phi_\mx\, \frac{\partial s}{\partial t} - K_\mx\,
{\rm div}_y\, \bigg\{\lambda_{\,\mx,w}(\bsvt_\mx) \nabla_y {\mathsf P} + \nabla_y \mathfrak{b}(\bsvt_\mx)
+ \widehat{\mathsf a}_{\mx,2}\,\lambda_{\,\mx,w}(\bsvt_\mx)\,
\nabla_y{\EuScript F}_{w}\left(\bsvt_\mx, \frac{\partial \bsvt_\mx}{\partial t}\right) \bigg\}
= 0;
\end{equation}
\begin{equation}
\label{fanta-2}
- \Phi_\mx\, \frac{\partial s}{\partial t}
- K_\mx\,
{\rm div}_y\, \bigg\{\lambda_{\,\mx,n}(\bsvt_\mx) \nabla_y {\mathsf P} -
\nabla_y \mathfrak{b}(\bsvt_\mx)
+ \widehat{\mathsf a}_{\mx,2}\,\lambda_{\,\mx,n}(\bsvt_\mx)\,
\nabla_y{\EuScript F}_{n}\left(\bsvt_\mx, \frac{\partial \bsvt_\mx}{\partial t}\right) \bigg\} = 0.
\end{equation}
We add the equations (\ref{fanta-1}) and (\ref{fanta-2}), to have:
\begin{small}
\begin{equation}
\label{fanta-3}
- {\rm div}_y \bigg\{\lambda_\mx(\bsvt_\mx) \nabla_y {\mathsf P}
+
\widehat{\mathsf a}_{\mx,2}
\left[\lambda_{\,\mx,w}(\bsvt_\mx)\,
\nabla_y{\EuScript F}_{w}\left(\bsvt_\mx, \frac{\partial \bsvt_\mx}{\partial t}\right)
+
\lambda_{\,\mx,n}(\bsvt_\mx)\,
\nabla_y{\EuScript F}_{n}\left(\bsvt_\mx, \frac{\partial \bsvt_\mx}{\partial t}\right)
\right]
\bigg\} = 0.
\end{equation}
\end{small}
Then we can impose the second condition on the functions ${\EuScript F}_{w}, {\EuScript F}_{n}$.
Namely, we set:
\begin{equation}
\label{fantik-1}
\lambda_{\,\mx,w}(\bsvt_\mx)\,
\nabla_y{\EuScript F}_{w}\left(\bsvt_\mx, \frac{\partial \bsvt_\mx}{\partial t}\right)
+
\lambda_{\,\mx,n}(\bsvt_\mx)\,
\nabla_y{\EuScript F}_{n}\left(\bsvt_\mx, \frac{\partial \bsvt_\mx}{\partial t}\right)
= 0.
\end{equation}

Now the simple calculations lead to the following result:

\begin{lemma}
\label{lem-deux-fonc-ff}
Let the functions ${\EuScript F}_{w}, {\EuScript F}_{n}$ satisfy the conditions
(\ref{F-ass-on-ff+1}) and (\ref{fantik-1}). Then
\begin{small}
\begin{equation}
\label{fantik-3}
\nabla_y{\EuScript F}_{w}\left(\bsvt_\mx, \frac{\partial \bsvt_\mx}{\partial t}\right) =
-
\frac{\lambda_{\,\mx,n}(\bsvt_\mx)}{\lambda_{\,\mx}(\bsvt_\mx)}\,
\nabla_y \frac{\partial \bsvt_\mx}{\partial t}
\quad {\rm and} \quad
\nabla_y{\EuScript F}_{n}\left(\bsvt_\mx, \frac{\partial \bsvt_\mx}{\partial t}\right) =
\frac{\lambda_{\,\mx,w}(\bsvt_\mx)}{\lambda_{\,\mx}(\bsvt_\mx)}\,
\nabla_y \frac{\partial \bsvt_\mx}{\partial t}.
\end{equation}
\end{small}
\end{lemma}

Lemma \ref{lem-deux-fonc-ff} implies that (\ref{fanta-3}) becomes:
$- {\rm div}_y \big\{\lambda_\mx(\bsvt_\mx) \nabla_y {\mathsf P} \big\} = 0$.
Then it follows from  (\ref{F-present-2})$_4$ that the boundary conditions for
the real saturation $s$ as well as for the function ${\mathsf P}$ on the interface
$\Gamma_{\fr\mx}$ (see the beginning of Section \ref{glob-mod-cras} for the definition of
$\Gamma_{\fr\mx}$) do not depend on the variable $y$. This fact allows us to prove the
following result (see Lemma 1 in \cite{Jurak} for similar arguments).
\smallskip
\begin{lemma}
\label{lem-pglob-ne-zavis-y}
The function ${\mathsf P}$ does not depend on the variable $y$, i.e.,
$\nabla_y {\mathsf P} = 0$ in $Y_\mx\times \Omega_T$.
\end{lemma}

Now, taking into account the results of Lemma \ref{lem-pglob-ne-zavis-y},
from equation (\ref{fanta-1}), (\ref{fantik-3}), and (\ref{F-present-2})$_3$, we obtain,
finally, the desired non-equilibrium imbibition equation. It reads:
\begin{equation}
\label{fantik-9}
\Phi_\mx\, \frac{\partial s}{\partial t} + K_\mx\,
{\rm div}_y\, \bigg\{\digamma(\bsvt_\mx)\, \nabla_y
p_c\left(\bsvt_\mx, \frac{\partial \bsvt_\mx}{\partial t}\right) \bigg\} = 0
\,\,\, {\rm in} \,\, Y_\mx\times\Omega_T,
\end{equation}
where
\begin{equation}
\label{fantik-8}
\digamma(\bsvt_\mx) \eqdef \frac{\lambda_{\,\mx,n}(\bsvt_\mx)\,
\lambda_{\,\mx,w}(\bsvt_\mx)}{\lambda_\mx(\bsvt_\mx)}.
\end{equation}
Thus the homogenized double porosity Kondaurov model contains the global equations
(\ref{F-present-1-sal}) coupled with the boundary value problem for the non-equilibrium
imbibition equation (\ref{fantik-9}).
\medskip

\begin{remark}
\label{on-imbib-rem}
Notice that in contrast to the classical case (see, e.g., \cite{Jurak} and the references therein)
or the case of the global Barenblatt model \cite{michel+ba+lp}, equation (\ref{fantik-9}) is
integro-differential. This fact shows explicitly the impact of the capillary non-equilibrium
on the mass exchange between the fracture system and blocks via the source terms
${\EuScript Q}_w, {\EuScript Q}_n$ in (\ref{F-present-1-sal}).
\end{remark}

\begin{remark}
\label{rem-limit-cas}
Notice that if $\tau_\mx = 0$ (equilibrium state) then, as it was shown in Remark
\ref{remark-on-hom-sys}, $\bsvt_\mx = s$ and, in addition, due to (\ref{const-aaa}),
$\widehat{\mathsf a}_{\mx,2} = 0$. Thus, equation (\ref{fantik-9}) becomes:
$$
\Phi_\mx\, \frac{\partial s}{\partial t} - K_\mx\,
\Delta_y\, \mathfrak{b}(s) = 0 \,\,\, {\rm in} \,\, Y_\mx\times\Omega_T,
$$
where the function $\mathfrak{b}$ is defined in (\ref{upsi-1}). This is exactly (with
evident modifications due to a special form of the capillary pressure $\pi_{\rm c}$)
the classical imbibition equation in the equilibrium case (see, e.g., formula (24) in
\cite{Jurak}).
\end{remark}

\section*{Concluding remarks}

In the framework of Kondaurov's formalism \cite{Kond}, a non-equilibrium porous medium
saturated by two fluids is described by a dependence of the thermodynamical potential on
a number of constitutive parameters. Using the relations which guarantee a thermodynamical model
consistency, it is possible to calculate the capillary pressure function and the right-hand
side of the kinetic equation. The first one determines the capillary driven fluxes
and the second one describes the capillary relaxation
processes. This approach has a number of advantages in contrast to Barenblatt's model
(see, e.g., \cite{bar-patz-sil}) whose application is restricted to weakly
non-equilibrium flows. Turning to the model considered in this Note, we observe that
in practice, the fractured porous medium is usually modeled by two-superimposed
continua, a connected fracture system and a system of topologically disconnected matrix blocks
(see, e.g., \cite{ak-lp-AA} and the reference therein). Therefore, we are facing a problem
of description of a highly heterogeneous medium, where each block is described
by Kondaurov's model. A distinctive feature of this model is as follows. It enables to take
into account the impact of the capillary non-equilibrium on the mass exchange between the
fissure system and the blocks. The analysis of the homogenized system carried out in this Note
shows some new aspects which are briefly discussed below. We focus our attention on two main points.
\medskip

\noindent {\bf (i)} {\bf Numerical aspects of Kondaurov's model.} From the mathematical
point of view, the double porosity models like (\ref{F-present-1-sal})-(\ref{F-present-2})
are rather complex systems of PDE involving $(2\d + 1)$ variables $(x, y, t)$ instead
of $(d + 1)$ for the initial mesoscopic system. However, we know (see, e.g., Ch. 10 in
\cite{hor}) that the double porosity model in contrast to the mesoscopic one does not
require the length scale of the block to be grid resolved. This enables us to solve
macroscopic problems numerically and\, justifies the importance of the homogenization
process in the study of non-equilibrium flows, like Kondaurov’s flow model, in
double porosity media. Our next step in this Note is to pass from the matrix problem
formulated in terms of phase pressures to the non-equilibrium
imbibition equation (\ref{fantik-9}). Evidently, the new formulation of the homogenized problem is
more easier for the numerical simulation because the number of the unknown functions and, consequently,
the standing equations is lower than for problem (\ref{F-present-1-sal})-(\ref{F-present-2}).
Notice that for the case of equilibrium two-phase flow in double porosity media (see, e.g.,
\cite{blm,hor,yeh2}) the numerical analysis of the global model can be done in two main ways.
The first one is to deal directly with the global model involving an equilibrium imbibition
equation, using the numerical resolution of this equation by the approach proposed in \cite{brez}.
The second one is the linearization of the non-linear equilibrium imbibition equation
in the sense of \cite{AMPP,Arb-simpl} or like in \cite{Jurak} for the case of the double
porosity media with thin fissures. In this case the homogenized system becomes fully homogenized
(i.e., does not involve any coupling with a matrix problem) and
the numerical simulation is much more easier without great loss of accuracy. Thus, our further work
is to generalize these approaches to the analysis of the global Kondaurov model
(\ref{F-present-1-sal})-(\ref{fantik-9}).
\medskip

\noindent{\bf (ii)} {\bf Mathematical analysis of Kondaurov's model.} As it was underlined in
\cite{ak-lp-AA}, we carry out our work with eye to a rigorous mathematical analysis of Kondaurov's
model. To this end, in Section \ref{phys-mod}, we define rigorously the capillary pressure and mobility
functions. The main results of the Note are given in Section \ref{sec-non-eq-imb}.
The key point here is the definition of non-equilibrium global pressure.
The global pressure function for degenerate (i.e., when the mobility functions vanish for
the wetting phase saturation taking the values zero or one) equilibrium multiphase flows
in porous media plays a crucial role, in particular, for the proof of compactness results.
This is also the case for the non-equilibrium two-phase flows like Kondaurov's flow model
or the Hassanizadeh model (see, e.g. \cite{hass}). It enables to apply the ideas of
\cite{wgh-two} in the proof of the existence result, including the proof of the maximum
principle for the real saturation. The notion of the non-equilibrium global pressure along
with the non-equilibrium matrix imbibition equation will play an important role in the
rigorous justification of the homogenization result obtained by formal asymptotic expansions
in \cite{ak-lp-AA}.

Thus the main novelty of the paper with respect to the existing literature, is the
introduction of the non-equilibrium global pressure and derivation of the non-equilibrium imbibition
equation. These results will allow us to continue the development of the theory of non-equilibrium
multiphase flows in porous media.

\section*{Acknowledgements}

This work was supported by the Russian Scientific Fund [grant number 15-11-00015].
The work of L. Pankratov was partially supported by the Russian Academic
Excellence Project {\sl 5top100}.

\renewcommand{\baselinestretch}{0.6}

\begin{small}

\end{small}


\begin{thebibliography}{99}

\bibitem{our-siam} B.~ Amaziane, S.~Antontsev, L.~Pankratov, A.~Piatnitski,
Homogenization of immiscible compressible two-phase flow in
porous media : application to gas migration in a nuclear waste
repository, {\it SIAM MMS},  {\bf 8}, (2010), 2023-2047.

\bibitem{AMPP} B.~Amaziane, J.~P.~Mili\v{s}i\' c, M.~Panfilov, L.~Pankratov,
Generalized nonequilibrium capillary relations for two-phase flow through
heterogeneous media, {\it Phys. Rev. E}, (2012) {\bf 85}, 016304, 1-18.

\bibitem{wgh-two} B. Amaziane, L. Pankratov, A. Piatnitski, The existence of
weak solutions to immiscible compressible two-phase flow in porous media:
the case of fields with different rock-types,
{\it Discrete Contin. Dyn. Syst. Ser. B} {\bf 18} (2013) 1217-1251.

\bibitem{m2as} B.~Amaziane, L.~Pankratov, Two-scale convergence of a model for water-gas flow through
double porosity media, M2AS, 2015, DOI: 10.1002/mma.3493.

\bibitem{michel+ba+lp} B.~Amaziane, M.~Panfilov, L.~Pankratov,
Homogenized model of two-phase ow with local non-equilibrium in double porosity media,
Submitted to {\it Journal of Mathematical Fluid Mechanics}, 2015.

\bibitem{ant-kaz-mon1990} S.~N. Antontsev, A.~V. Kazhikhov and V.~N. Monakhov,
{\it Boundary Value Problems in Mechanics of Nonhomogeneous Fluids}, North-Holland,
Amsterdam, 1990.


\bibitem{Arb-simpl} T. Arbogast, A simplified dual-porosity model for two-phase flow,
in Computational Methods in Water Resources IX, Vol. 2 (Denver, CO, 1992):
Mathematical Modeling in Water Resources, T.F. Russell, R.E. Ewing, C.A. Brebbia, W.G. Gray, and
G.F. Pindar, eds., Comput. Mech., Southampton, U.K., 1992, pp. 419-426.

\bibitem{BAKH-PAN} N. Bakhvalov, G. Panasenko, Homogenisation: averaging processes
in periodic media, Dordrecht: Kluwer Academic Publishers, 1989.

\bibitem{bar-patz-sil} G.~I.~Barenblatt, T.~W.~Patzek, and D.~B.~Silin,
The mathematical model of non-equilibrium effects in water-oil displacement,
{\it SPE Journal}, {\bf 8}:4, (2003), 409-416.

\bibitem{BLP} A.~Bensoussan, J.L.~Lions, G.~Papanicolaou,
{\it Asymptotic analysis for periodic structures}, North-Holland, Amsterdam, 1978.

\bibitem{bot} S.~Bottero, S.M.~Hassanizadeh, P.J.~Kleingeld, T.~Heimovaara,
Nonequilibrium capillarity effects in two-phase flow through porous media at different
scales, {\it Water Resour. Res.}, {\bf 47}, (2011).

\bibitem{blm} A.~Bourgeat, S.~Luckhaus, A.~Mikeli\'c, Convergence of the
homogenization process for a double-porosity model of immicible two-phase flow,
{\it SIAM J. Math. Anal.}, {\bf 27}, (1996), 1520-1543.

\bibitem{Bourgeat-Panfilov} A.~Bourgeat, M.~Panfilov, Effective two-phase flow
through highly heterogeneous porous media: capillary nonequilibrium effects,
{\it Computational Geosciences} {\bf 2}, (1998), 191-215.

\bibitem{brez} A.~E.~Berger, H.~Brezis, J.~C.~ W.~Rogers, A numerical method for
solving the problem $u_t - \Delta f(u) = 0$, {\it RAIRO - Analyse num\'erique},
{\bf 13}:4, (1979), 297-312.

\bibitem{pop} X.~Cao, I.S.~Pop, Two-phase porous media flows with dynamic capillary effects
and hysteresis: Uniqueness of weak solutions, {\it Computers and Mathematics with
Applications}, {\bf 69}, (2015), 688-695.

\bibitem{GC-JJ} G.~Chavent, J.~Jaffr\'e, {\it Mathematical Models and Finite
Elements for Reservoir Simulation}, North-Holland, Amsterdam, 1986.

\bibitem{coussy} O.~Coussy, {\it Poromechanics}, Wiley, New-York, 2004.

\bibitem{gals-cras} C. Galusinski and M. Saad, Weak solutions for immiscible
compressible multifluid flows in porous media, {\it C. R. Acad. Sci. Paris, S\'er. I},
{\bf 347}, (2009), 249-254.

\bibitem{hor} U.~Hornung, Homogenization and porous media, Springer-Verlag, New York, 1997.

\bibitem{Jurak} M.~Jurak, L.~Pankratov, A.~Vrba\v{s}ki, A fully homogenized model for
incompressible two-phase flow in double porosity media, {\it Applicable Analysis} (2015),
DOI: 10.1080/00036811.2015.1031221.

\bibitem{hass} S.~ M.~Hassanizadeh, W.~G.~Gray, Thermodynamic basis of capillary
pressure in porous media, {\it Water resources research}, {\bf 29}:10 (1993), 3389-3405.

\bibitem{KRS} J.~Koch, A.~ R\"atz, B. Schweizer, Two-phase flow equations with a dynamic
capillary pressure, {\it Eur. J. Appl. Math.}, {\bf 24}:1, (2013), 49-75.

\bibitem{Kond} V.~I.~Kondaurov, A non-equilibrium model of a porous medium saturated with
immiscible fluids, {\it Journal of Applied Mathematics and Mechanics}, {\bf 73} (2009), 88-102.

\bibitem{a-kon} A.~Konyukhov, A.~Tarakanov, On two approaches in investigation of
non-equilibrium effects of filtration in a porous medium, in {\it Proceedings of
the Fifth Biot Conference on Poromechanics}, ASCE 2013, 2307-2316.

\bibitem{ak-lp-AA} A.~Konyukhov, L.~Pankratov, Upscaling of an immiscible non-equilibrium
two-phase flow in double porosity media, {\it Applicable Analysis} (2015),
DOI 10.1080/00036811.2015.1064524.

\bibitem{panf} M.~Panfilov, {\it Macroscale models of flow through highly
heterogeneous porous media}, Kluwer Academic Publishers, London, 2000.

\bibitem{richard} J.~G.~Richardson, {\it Flow through porous media}, in:
in Handbook of Fluiddynamics, Ed. by V. L. Streeter, McGraw Hill, 16-65, 1961.

\bibitem{sal-bru-2010a} H.~Salimi, J.~Bruining, Improved prediction of oil
recovery from waterflooded fractured reservoirs using homogenization
{\it SPE Reserv. Evalu. Eng.}, {\bf 13}:1 (2010), 44-55.

\bibitem{sal-bru} H.~Salimi, J.~Bruining, Upscaling in vertically fractured oil reservoirs
using homogenization, {\it Transport in porous media}, {\bf 84} (2010), 21-53.

\bibitem{salimi} H.~ Salimi, J.~Bruining, Upscaling of fractured oil reservoirs
using homogenization including non-equilibrium capillary pressure and relative permeability,
{\it Computational Geoscience} {\bf 16} (2012), 367-389.

\bibitem{SP} E.~Sanchez-Palencia, {\it Non-homogeneous media and vibration theory},
Springer-Verlag, Berlin, 1980.

\bibitem{das-book} {\it Upscaling Multiphase Flow in Porous Media} (ed. by D.B. Das and
S.M. Hassanizadeh), Springer, Dordrecht, The Netherlands, 2005.

\bibitem{yeh2} L.~M.~Yeh, Homogenization of two-phase flow in fractured media,
{\it Math. Methods Appl. Sci.} {\bf 16} (2006), 1627-1651.


\end{thebibliography}
\end{document}